\documentclass[journal,10pt]{IEEEtran}
\IEEEoverridecommandlockouts
\usepackage{cite}
\usepackage{amsmath,amssymb,amsfonts}
\usepackage{subfigure}
\usepackage{graphicx,epsfig,balance}
\usepackage{algorithmic}
\usepackage{graphicx}
\usepackage{textcomp}
\usepackage{xcolor}
\usepackage{booktabs}
\def\BibTeX{{\rm B\kern-.05em{\sc i\kern-.025em b}\kern-.08em
    T\kern-.1667em\lower.7ex\hbox{E}\kern-.125emX}}
\begin{document}

\title{A Real-World Radio Frequency Signal Dataset Based on LTE System and Variable Channels}

\author{Shupeng Zhang$^\dag$, Yibin Zhang$^\dag$, Xixi Zhang$^\dag$, Jinlong Sun$^\dag$,
 Yun Lin$^{\star}$, Haris Gacanin$^{\ddag}$, \\ Fumiyuki Adachi$^{\dag\dag}$, and Guan Gui$^\dag$\\
~\\

$^\dag$College of Telecommunications and Information Engineering, NJUPT, Nanjing, China\\
$^\star$College of Information and Communication Engineering, Harbin Engineering University, Harbin, China\\
$^{\ddag}$Institute for Communication Technologies and Embedded Systems, RWTH Aachen University, Aachen, Germany\\
$^{\dag\dag}$International Research Institute of Disaster Science (IRIDeS), Tohoku University, Sendai, Japan\\
$^{*}$E-mails: harisg@ice.rwth-aachen.de, fumiyuki.adachi.b4@tohoku.ac.jp, guiguan@njupt.edu.cn
}

\maketitle

\begin{abstract}
%Considering continuously updating of attacks and intrusions, the security of wireless communication system has became one of the key stages gradually over the last decade.
Radio Frequency Fingerprint (RFF) identification on account of deep learning has the potential to enhance the security performance of wireless networks. Recently, several RFF datasets were proposed to satisfy requirements of large-scale datasets. However, most of these datasets are collected from 2.4G WiFi devices and through similar channel environments. Meanwhile, they only provided receiving data collected by the specific equipment. This paper utilizes software radio peripheral as a dataset generating platform. Therefore, the user can customize the parameters of the dataset, such as frequency band, modulation mode, antenna gain, and so on. In addition, the proposed dataset is generated through various and complex channel environments, which aims to better characterize the radio frequency signals in the real world. We collect the dataset at transmitters and receivers to simulate a real-world RFF dataset based on the long-term evolution (LTE). Furthermore, we verify the dataset and confirm its reliability. The dataset and reproducible code of this paper can be downloaded from GitHub\footnote {GitHub link: https://github.com/njuptzsp/XSRPdataset}.
\end{abstract}


\section{Introduction}
Recently, as wireless communication keeps advancing, the application of Internet of Things (IoT) has become an indispensable part of daily life. Meanwhile, cyber attack like impersonation attack happens more often. The hacker is able to duplicate the identity information to spoof devices \cite{ZhaoWCL2021,Abducyber2018,Lia2019,Anbias2020}. Therefore, the security of wireless communication becomes a vital issue \cite{Jinproblem2021}. Radio Frequency Fingerprinting (RFF) is regarded as a technology to identify wireless devices and to strengthen the security of wireless communication \cite{Denga2017}. The RFF is the feature caused by the hardware imperfections of analog circuitry. These hardware imperfections are generated during the production process. Therefore, RFF is unique and applicable to devices identification\cite{Yangradio2021,NouichiIot2019,Sola2020}.

\begin{figure}[t]
	\centering{\includegraphics[width=8.8cm,height=2.8cm]{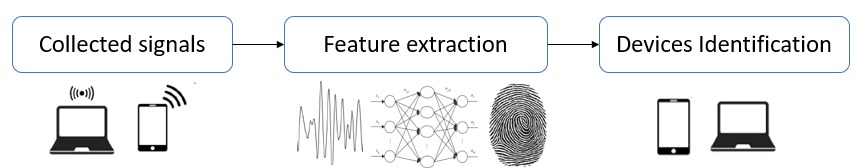}}
	\caption{Devices identification using RFF.}
	\label{system}
\end{figure}

Recognizing the most relevant functions from a mixture containing large numbers of signal radios is a challenge\cite{WangJASC2021,WangWCL2021,QinWC2019,ZhangTCCN2022}. It is protocol dependent, therefore it needs domain knowledge and advanced testing facility. On the contrary, methods based on neural network are gaining appeal because they can automatically identify features with considering protocol: only raw in-phase (I) and quadrature (Q) components of the RF signal samples are sufficient for recognition, which greatly facilitates application of RFF identification. Hence, many researchers work to apply deep learning to RFF identification \cite{YuA2019,Sanno2019,Zonga2020,Liur2021,PengWCL2022,YuWCL2021,YuCL2022}. The main stages in the RFF identification are shown in Fig.~\ref{system}. Significantly, this technology requires high-quality datasets. Recently, several datasets were proposed. They are shown as follows.
\begin{itemize}
	\item ORACLE: The ORACLE dataset recorded raw IQ symbols collected from transmissions of 16 USRP X310 transmitter signals and demodulated IQ samples collected after equalizing transmissions of 16 IQ imbalance configurations\cite{Sanoracle2019}.
	\item The dataset recorded bluetooth signals from 27 various phones and changed sampling rate in each collection. The data acquisition system for creating dataset was described in detail. Then, the reliability of two methods based on transient bluetooth signals were tested by the dataset\cite{Ezua2020}.
	\item ADS-B: A large scale radio dataset is created, which is dependent on a peculiar aviation system, i.e., Automatic Dependent Surveillance-Broadcast (ADS-B). This dataset captured ADS-B signals outdoors and labelled them. Then, they cleaned and sorted the data and created a high-quality dataset finally\cite{Tularge2021}.
\end{itemize}

However, the above datasets are signals collected from 2.4G wireless devices such as smartphones and laptops. The signal frequency band is single and not universal. Also, some of them are collected under similar ideal wireless channels. Except these, datasets described above only recorded signals data at the receiver-side. Therefore, we use a wireless communication system based on LTE \cite{Labibana2015,Yinlte2021} and generate 800 MHz signals. We transmit signals through different real wireless channels collecting both the transmission and reception data in a 800 MHz LTE signal frequency band where the one can simulate a more realistic communication environment. Although, the channel environment does not affect RFF, it influences the received signal and RFF feature extraction\cite{Soltanmore2020}. Our main contributions of the letter are shown below.
\begin{itemize}
\item To meet the requirements of large-scale dataset, we use a wireless communication system based on LTE and generate a dataset by using 10 different XSRP as transmitters and storing transmission and reception data.
\item To solve the problem of single frequency band and lack of complex channel environment, we set emission frequency as 800 MHz and generate two channel environments including Line of Sight (LOS) channel and Non Line of Sight (NLOS) channel\cite{Wanglos2019,Aonoperformance2021}.
\item We validate the dataset by using simple CNN. Judging from the classification accuracy, the dataset is reliable and can be used in deep learning-based RFF identification techniques.
\end{itemize}

\section{System Model}
\subsection{Wireless Communication System Based on LTE}
In this part, we introduce the wireless communication system based on LTE. LTE provides better
coverage, improved system capacity, and signals with high peak value in a highly efficient manner. The signal source data is generated on the computer and carried out a series of processing based on LTE. Then, the signal data are sent to the XSRP as transmitter and transmitted through wifeless channel to another XSRP as receiver. As shown in Fig. \ref{LTE}, the detailed stages of wireless communication system based on LTE are as follows:

\subsubsection{Generate source}
We use 10 different XSRP as transmitters to meet the need of large-scale dataset. This means we need enough radio source data to process and transmit. If we transmit same signal data every time, the dataset is unreasonable. We should ensure that most of radio signal source is different. Therefore, we use Matlab to PCM encode an 18 minute MP3 file and then generate five thousands rows of source data. We select 400 rows randomly as radio source data in every transmission.

\subsubsection{Channel code}
To enhance data transmission efficiency and decrease bit error rate, the channel coding type is 1/2 convolutional code. That is, two bits are output for each input bit. We represent the data after channel coding as $x(t)$.

\subsubsection{Modulation mapping}
The modulation type is 16 Quadrature Amplitude Modulation (QAM). We map the data to the constellation according to the 16QAM mapping table. Modulation mapping is to improve signal anti-interference ability and facilitate wireless communication.

\subsubsection{Generate pilot}
We generate 2 pilots for the channel estimation at the receiver-side.

\subsubsection{Resource mapping}
This is to map the digital signal of 01 to the specific time and frequency of electromagnetic wave.

\subsubsection{IFFT}
Inverse Fast Fourier Transform (IFFT) is to transfer frequency domain data to time domain.

\subsubsection{Add CP}
Add CP is to eliminate Inter-Carrier Interference (ICI) and Inter-Symbol Interference (ISI) and it will be described later. Finally, we perform DA conversion and transmit the analog signal. The RF signal after up conversion is shown as:
\begin{equation}
X(t)=s(t) e^{-j 2 \pi f_{c}^{T x} t},
\end{equation}
where $s(t)$ represents the baseband signal; $f_{c}^{T x}$ is the carrier frequency of the transmitter.

After processing in the XSRP as the transmitter, the radio signals carry RFF. The rest of the processing procedures at receiver-side will not be introduced because they are all restoration in the previous stage. Considering that wireless channels affect reception, the received signal is $R(t)=\alpha_{c} X(t)$, where $\alpha_{c}$ is the wireless channel fading gain. The baseband signal after down conversion is represented as:
\begin{equation}
\label{eq2}
\begin{aligned}
Y(t) &=R(t) e^{j 2 \pi f_{c}^{R x} t+\phi} \\
&=\alpha_{c} s(t) e^{j 2 \pi \Delta f t+\phi},
\end{aligned}
\end{equation}
where $f_{c}^{R x}$ is the carrier frequency of receiver, $\phi$ is the receiver phase error, $\Delta f=f_{c}^{R x}-f_{c}^{T x}$ is the frequency offset between transmitter and receiver. The phase rotation factor $e^{j 2 \pi \Delta f t+\phi}$ in (\ref{eq2}) changes with time $t$.

\begin{figure*}[htbp]
	\centering{\includegraphics[width=6.0 in]{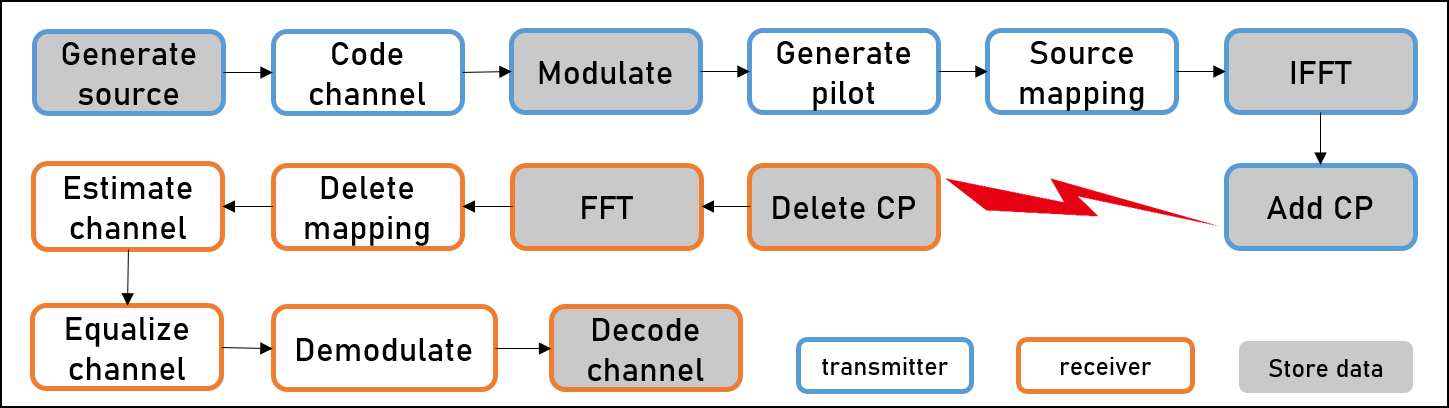}}
	\caption{Wireless communication system based on LTE.}
	\label{LTE}
\end{figure*}

Main system parameters are shown as Tab.~\ref{parameter}. In this procedure, the data collection and processing are all completed by computers. We can choose any part of the system's data to store freely. For the purpose of RFF identification and follow-up extended experiment, we choose some parts of the system's data to store in the computer. All above mentioned communication modules are all shown in Fig.~\ref{LTE}.

\begin{table}[b]
	\centering
	\caption{System parameters and default values.}
	\begin{tabular}{c|c}
		\toprule
		{\bf System parameter} & {\bf Default setting} \\ \hline
		Modulation type & 16QAM \\ \hline
		Channel coding type & 1/2 convolutional code \\ \hline
		SNR & 20 dB \\ \hline
		Transmission frequency & 800 MHz\\ \hline
		Reception frequency & 800 MHz\\
		\bottomrule
	\end{tabular}
    \label{parameter}
\end{table}

\subsection{LTE Frame Structure}
\begin{figure*}[htbp]
	\centering{\includegraphics[width=6.3 in]{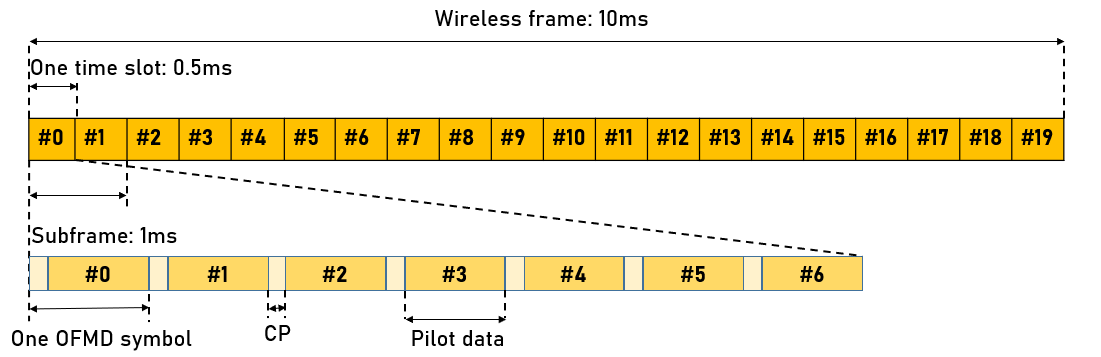}}
	\caption{LTE frame structure.}
	\label{frame}
\end{figure*}

In this part, we describe LTE frame structure. From Fig.~\ref{frame}, we can see that one 10 ms wireless frame contains ten 1 ms subframes. Each subframe consists of two 0.5 ms time slots. Each time slot consists of 7 OFDM symbols. Useful data accounts for 6 symbols. Pilot data occupies symbol position 3. The CP is to copy the part within the last certain length of the OFDM signal and put it in the head of the OFDM signal. The OFDM signal that becomes longer after adding the CP is used as a new OFDM signal. In this way, ICI and ISI can be completely eliminated if the CP length is longer than the largest propagation time delay \cite{Noueva2017,Nacyc2015}.

\section{Dataset Generation Approach}
\label{sec3}

\subsection{Channel Environment}
\begin{figure}[htbp]
	\centering{\includegraphics[width=3.3 in]{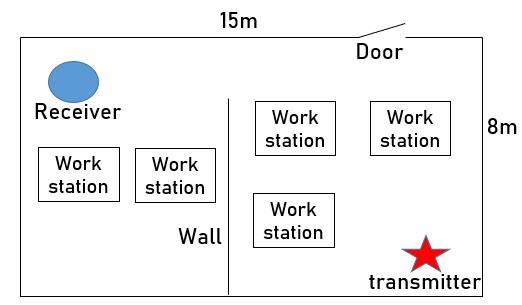}}
	\caption{NLOS channel environment constructed in laboratory.}
	\label{NLOS}
\end{figure}

\begin{figure}[htbp]
	\centering{\includegraphics[width=3.3 in]{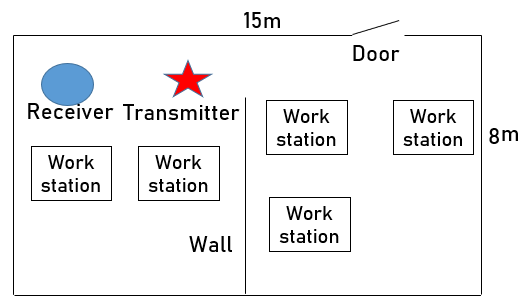}}
	\caption{LOS channel environment constructed in laboratory.}
	\label{LOS}
\end{figure}

We select our laboratory as the place to transmit and receive signals. Due to many people, personal computers, work station and closed room, our laboratory has complex electromagnetic environment. On this basis, we design two kinds of channel environments including: NLOS and LOS channel environments. They are shown in Fig.~\ref{NLOS} and Fig.~\ref{LOS}. In the situation of NLOS, wireless signals are received in many ways containing reflection, scattering and diffraction, and multipath effect will bring many troubles like time delay synchronization, link instability and so on. In NLOS channel environment, the distance between transmitter and receiver is far and there will be people walking around sometimes in our laboratory, hence the channel conditions are changing. This can help us simulate more channel conditions and SNR regimes. Under the condition of LOS, the wireless signal propagates in a straight line between the transmitter and the receiver without shielding. This is an ideal channel environment and the distance between the transmitter and the receiver is close.

\subsection{Hardware Setup}
\begin{figure}[htbp]
	\centering{\includegraphics[width=3.5 in]{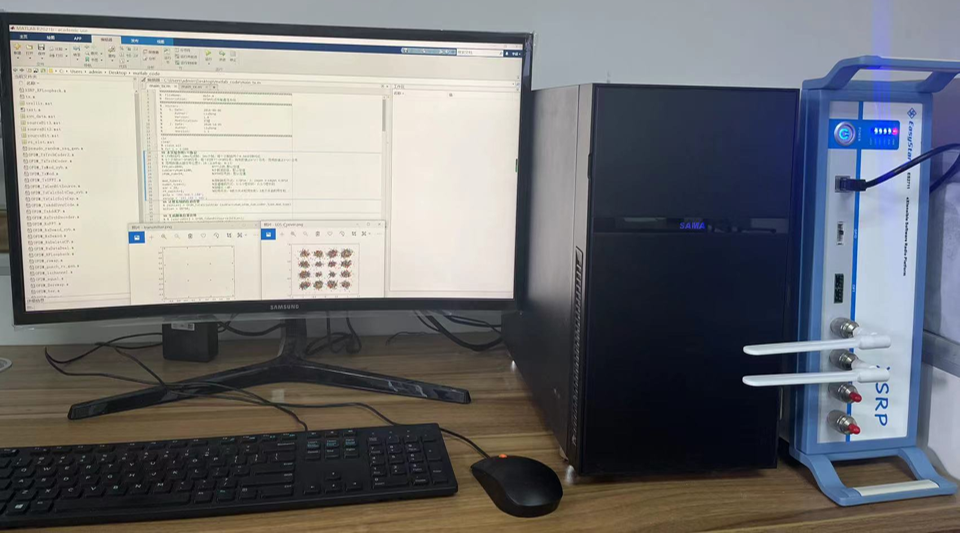}}
	\caption{Hardware setup.}
	\label{hardware}
\end{figure}

The hardware configuration of wireless communication system is represented below:
 \begin{enumerate}
 	\item  Signal receiving equipment: We use XSRP as transmitter and receiver. XSRP is a software defined radio platform. We connect it to the computer with network cable and it can make an ordinary computer work like high broadband software radio equipment. In essence, it acts as the digital baseband and intermediate frequency part of the communication system.
 	\item Signal processing equipment: As is shown in Fig.~\ref{hardware}, there is a computer at each transmitter-side and receiver-side which is used for signal processing and storage.
    \item Computer configuration: CPU is Intel(R) Core(TM) i7-10700 @ 2.90 GHz, Random Access Memory is 16 GB, and Solid State Disk is 1 TB.
 \end{enumerate}

The dataset consists of two parts including: transmitter-side data and receiver-side data. We store source data, modulation data, IFFT data and transmission data at transmitter-side. At receiver-side data, we store the data corresponding to the transmitter. We use five transmitters for every channel environment. Each transmission and reception last about 4 hours and collect 400 samples. The length of each sample is 28796. After storing ten times of transmission and reception using 10 different XSRP, we combine the stored data into dataset.

\section{Dataset Analysis and Verification}
\label{sec4}
In section \ref{sec3}, we introduced how to generate dataset. This section, we mainly analyze the dataset including the effect of different channel environments to the radio reception. Meanwhile, the reliability of the dataset is verified.

\begin{figure}[htbp]
  \centering
  \subfigure[{transmitted constellation}]{
   \includegraphics[width=1.65 in] {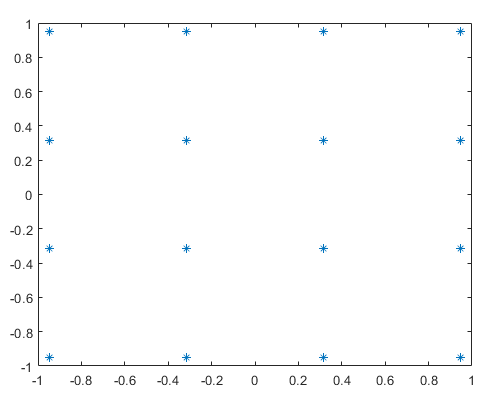}}
  \subfigure[received constellation]{
   \includegraphics[width=1.655 in] {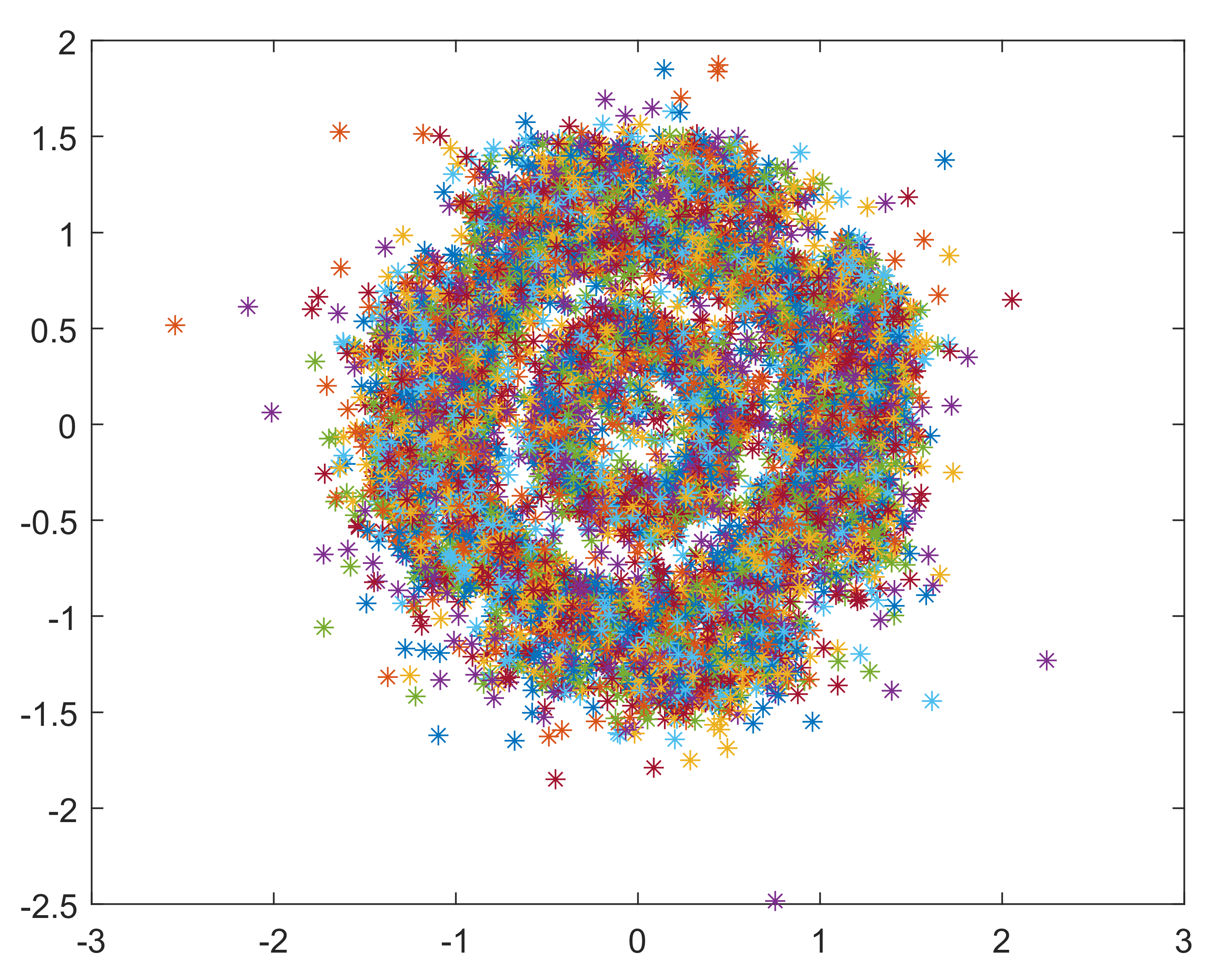}}
 \caption{Constellations of a transmitter and a receiver under NLOS channel environment.}
\label{bad}
\end{figure}

\begin{figure}[htbp]
  \centering
  \subfigure[{transmitted constellation}]{
   \includegraphics[width=1.65 in] {transmitterconstellation.png}}
  \subfigure[received constellation]{
   \includegraphics[width=1.655 in] {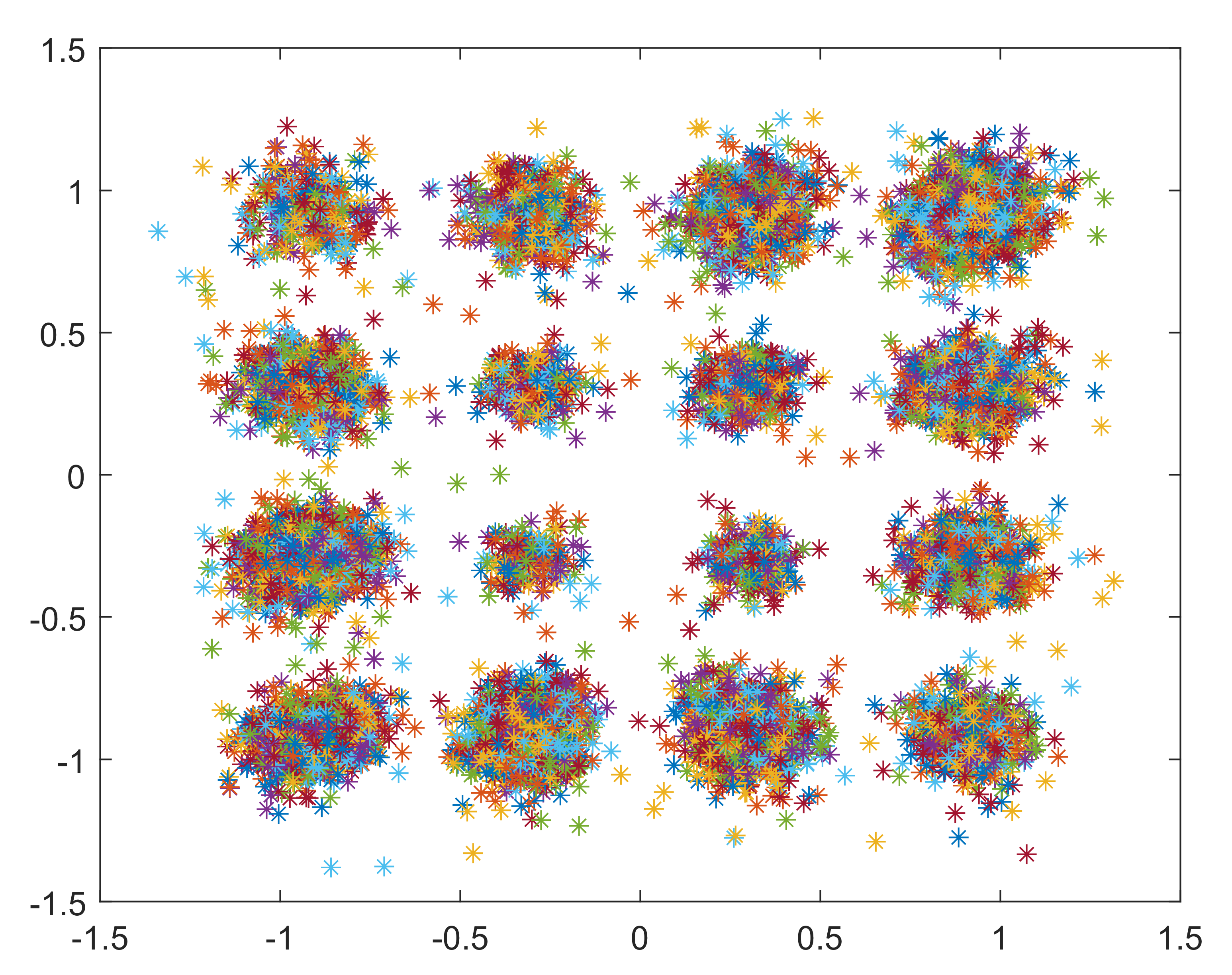}}
 \caption{Constellations of a transmitter and a receiver under LOS channel environment.}
\label{good}
\end{figure}

As is shown in Fig.~\ref{bad} and Fig.~\ref{good}, these two pictures are reviver constellation of 16QAM signals through different channel environments. From the comparison of the two pictures, NLOS channel environment lead to poor reception effect and it is difficult to get shape of the original constellation. Instead, transmitter and receiver constellation are similar in the situation of LOS channel environment. Therefore, the LOS channels often lead to ideal reception in our experiment. From the RFF introduced above, no matter poor or ideal reception effect, it will not affect the reliability of dataset and features extraction. Because RFF represents the physical feature due to hardware imperfections, it will not be changed by the reception effect. However, wireless channel is the main factor leading to the decline of accuracy in RFF identification based on deep learning. If a training set is collected in a specific channel environment, the neural network will eventually learn the fingerprint mixed with channel distortion, rather than pure inherent fingerprint. Therefore, we have collected the dataset which contains different kinds of channel environments and using this dataset to train the neural network can increase the robustness of the network.

\begin{figure}[htbp]
	\centering{\includegraphics[scale=0.6]{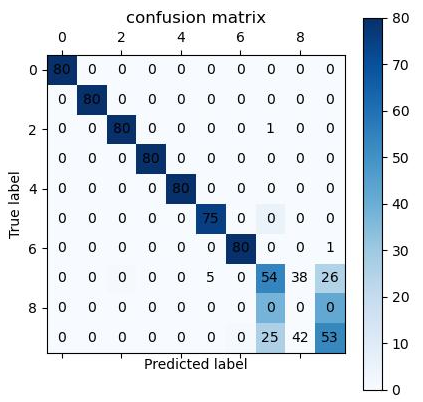}}
	\caption{Classification results with CNN.}
	\label{rx10}
\end{figure}

We also use a simple CNN to verify the reliability of dataset. CNN has the advantage of translation invariance: the converted weights detect the patterns at any position of each layer, and the signal is transmitted to the higher layer through the maximum pool layer. This feature enables the network to detect the patterns well. The dataset is divided into three portions, i.e., the training set, the validation set, and the test set. They account for 60\%, 20\%, and 20\% of the whole dataset respectively. The dataset has 4000 samples in total. Hence, there are 2400 samples in the training set, 800 samples in the validation set and 800 samples in the test set. To reduce training time and enhance performance, the epoch is set as 200 and batch size is set as 16. Also, an early stop mechanism was established to save time. The performance of the CNN model is shown as Fig.~\ref{rx10}. The classification accuracy is not very good but some species are classified very precisely. Hence, the dataset is reliable and more ways to improve classification accuracy of the dataset are worth exploring.

%\begin{figure}[htbp]
%	\centering{\includegraphics[width=3.5 in]{figure/badreception.png}}
%	\caption{NLOS channel environment reception.}
%	\label{bad}
%\end{figure}

%\begin{figure}[htbp]
%	\centering{\includegraphics[width=3.5 in]{figure/goodreception.png}}
%	\caption{LOS channel environment reception.}
%	\label{good}
%\end{figure}

\section{Conclusion}
In this paper, a real-world RFF dataset considering LTE and variable channels is proposed. We also present the main part about how to generate the dataset including system model and hardware setup. The test result using the dataset is also presented. As shown in the result, although the current results are not ideal, our dataset is confirmed that it can be used to deep learning. Finally, the dataset and matlab code is uploaded to Github. We expect that it is able to encourage more and more investigators to devote to the domains and propose more excellent methods to enhance the identification performance.

\section*{Acknowledgement}
This work was supported by the National Key Research and Development Program of China under Grant No. 2021ZD0113003, National Natural Science Foundation of China under Grant Number 61901228, China Postdoctoral Science Foundation Project under Grant Number 2021M702466.


\begin{thebibliography}{99}
\bibitem{ZhaoWCL2021}
R. Zhao, J. Yin, \emph{et al}., ``An efficient intrusion detection method based on dynamic autoencoder,'' \emph{IEEE Wireless Communications Letters}, vol. 10, no. 8, pp. 1707--1711, Aug. 2021.

\bibitem{Abducyber2018}
%M. S. Abdullah, A. Zainal, M. A. Maarof, and M. Nizam Kassim, ``Cyber-attack ffeatures for detecting cyber threat incidents from online news,'' in \emph{2018 Cyber Resilience Conference (CRC)}, Putrajaya, Malaysia, 2018, pp. 1--4.
S. I. Popoola, B. Adebisi, M. Hammoudeh, G. Gui, and H. Gacanin, ``Hybrid deep learning for botnet attack detection in the internet of things networks,'' \emph{IEEE Internet of Things Journal}, vol. 8, no. 6, pp. 4944--4956, Jun. 2021.

\bibitem{Lia2019}
%F. Li, X. Yan, Y. Xie, Z. Sang, and X. Yuan, ``A review of cyber-attack methods in cyber-physical power system,'' in \emph{2019 IEEE 8th International Conference on Advanced Power System Automation and Protection (APAP)}, Xi'an, China, 2019, pp. 1335--1339.
Y. Liu, Z. Su, and Y. Wang, ``Energy-efficient and physical-layer secure computation offloading in blockchain-empowered internet of things,'' \emph{IEEE Internet of Things Journal}, doi: 10.1109/JIOT.2022.3159248.

\bibitem{Anbias2020}
D. Antonioli, N. O. Tippenhauer, and K. Rasmussen, ``BIAS: bluetooth impersonation attacks,'' in \emph{2020 IEEE Symposium on Security and Privacy (SP)}, San Francisco, CA, USA, 2020, pp. 549--562.

\bibitem{Jinproblem2021}
L. Jin, X. Hu, Y. Lou, Z. Zhong, X. Sun, H. Wang, and J. Wu, ``Introduction to wireless endogenous security and safety: problems, attributes, structures and functions,'' \emph{China Communications}, vol. 18, no. 9, pp. 88--99, Sept. 2021.

\bibitem{Denga2017}
S. Deng, Z. Huang, and X. Wang, ``A novel specific emitter identification method based on radio frequency fingerprints,'' in \emph{2017 2nd IEEE International Conference on Computational Intelligence and Applications (ICCIA)}, Beijing, China, 2017, pp. 368--371.

\bibitem{Yangradio2021}
Y. Yang and T. Yan, ``Radio frequency fingerprint recognition method based on generative adversarial net,'' in \emph{2021 13th International Conference on Communication Software and Networks (ICCSN)}, Chongqing, China, 2021, pp. 361--364.

\bibitem{NouichiIot2019}
D. Nouichi, M. Abdelsalam, Q. Nasir, and S. Abbas, ``IoT devices security using RF fingerprinting,'' in \emph{2019 Advances in Science and Engineering Technology International Conferences (ASET)}, Dubai, United Arab Emirates, 2019, pp. 1--7.

\bibitem{Sola2020}
N. Soltanieh, Y. Norouzi, Y. Yang, and N. C. Karmakar, ``A review of radio frequency fingerprinting techniques,''" \emph{IEEE Journal of Radio Frequency Identification}, vol. 4, no. 3, pp. 222--233, Sept. 2020.

\bibitem{WangJASC2021}
Y. Wang, G. Gui, H. Gacanin, T. Ohtsuki, O. A. Dobre, and H. V. Poor, ``An efficient specific emitter identification method based on complex-valued neural networks and network compression,'' \emph{IEEE Journal on Selected Areas in Communications}, vol. 39, no. 8, pp. 2305--2317, Aug. 2021.

\bibitem{WangWCL2021}
P. Wang, Y. Cheng, B. Dong, and G. Gui, ``Binary neural networks for wireless interference identification,'' \emph{IEEE Wireless Commuications Letters}, vol. 11, no. 1, pp. 23--27, Jan. 2022.

\bibitem{QinWC2019}
Z. Qin, H. Ye, G. Y. Li, B.-H. F. Juang, ``Deep learning in physical layer communications,'' \emph{IEEE Wireless Communications}, vol. 26, no. 2, pp. 93--99, Feb. 2019.


\bibitem{ZhangTCCN2022}
X. X. Zhang, H. T. Zhao, \emph{et al.}, ``NAS-AMR: neural architecture search based automatic modulation recognition method for integrating sensing and communication system,'' \emph{IEEE Transactions on Cognitive Communications and Networking}, early access, doi: 10.1109/TCCN.2022.3169740.

\bibitem{Zonga2020}
L. Zong, C. Xu, and H. Yuan, ``A RF fingerprint recognition method based on deeply convolutional neural network,'' in \emph{2020 IEEE 5th Information Technology and Mechatronics Engineering Conference (ITOEC)}, Chongqing, China, 2020, pp. 1778--1781.

\bibitem{Liur2021}
D. Liu, M. Wang, and H. Wang, ``RF fingerprint recognition based on spectrum waterfall diagram,'' in \emph{2021 18th International Computer Conference on Wavelet Active Media Technology and Information Processing (ICCWAMTIP)}, Chengdu, China, 2021, pp. 613--616.

\bibitem{PengWCL2022}
Y. Peng, P. Liu, Y. Wang, G. Gui, B. Adebisi, and H. Gacanin, ``Radio frequency fingerprint identification based on slice integration cooperation and heat constellation trace figure,'' \emph{IEEE Wireless Commuications Letters}, vol. 11, no. 3, pp. 543--547, Mar. 2022.

\bibitem{YuWCL2021}
Y. Yang, A. Hu, Y. Xing, J. Yu, Z. Zhang, ``A data-independent radio frequency fingerprint extraction scheme,'' \emph{IEEE Wireless Commuications Letters}, vol. 11, no. 3, pp. 2524--2527, Nov. 2021.

\bibitem{YuCL2022}
Z. Yang, A. Hu, W. Xu, J. Yu, annd Y. Yang, ``An artificial radio frequency fingerprint embedding scheme for device identification,'' \emph{IEEE Commuications Letters}, vol. 26, no. 5, pp. 974--978, May 2021.

\bibitem{YuA2019}
J. Yu, A. Hu, G. Li, and L. Peng, ``A robust RF fingerprinting approach using multisampling convolutional neural network,'' \emph{IEEE Internet of Things Journal}, vol. 6, no. 4, pp. 6786--6799, 2019.

\bibitem{Sanno2019}
K. Sankhe, M. Belgiovine, \emph{et al.}, ``No radio left behind: radio fingerprinting through deep learning of physical-layer hardware impairments,'' \emph{IEEE Transactions on Cognitive Communications and Networking}, vol. 6, no. 1, pp. 165--178, 2019.

\bibitem{Sanoracle2019}
K. Sankhe, M. Belgiovine, \emph{et al.}, ``ORACLE: optimized radio classification through convolutional neuraL networks,'' in \emph{IEEE INFOCOM 2019 IEEE Conference on Computer Communications}, Paris, France, 2019, pp. 370--378.

\bibitem{Ezua2020}
E. Uzundurukan, Y. Dalveren, and A. Kara, ``A database for radio frequency fringerprinting of bluetooth Devices,'' \emph{Data}, vol. 5, no. 2, Jun. 2020.

\bibitem{Tularge2021}
Y. Tu, Y. Lin, H. Zha, J. Wang, and Y. Wang, ``Large-scale real-world radio signal recognition with deep learning,'' \emph{Chinese Journal of Aeronautics}, early access, doi: 10.1016/j.cja.2021.08.016.

\bibitem{Labibana2015}
M. Labib, V. Marojevic, and J. H. Reed, ``Analyzing and enhancing the resilience of LTE/LTE-A systems to RF spoofing,'' in \emph{2015 IEEE Conference on Standards for Communications and Networking (CSCN)}, Tokyo, Japan, 2015, pp. 315--320.

\bibitem{Yinlte2021}
P. Yin, L. Peng, J. Zhang, M. Liu, H. Fu, and A. Hu, ``LTE device identification based on RF fingerprint with multi-channel convolutional neural network,'' in \emph{2021 IEEE Global Communications Conference (GLOBECOM)}, Madrid, Spain, 2021, pp. 1--6.

\bibitem{Soltanmore2020}
N. Soltani, K. Sankhe, J. Dy, S. Ioannidis, and K. Chowdhury, ``More is better: data augmentation for channel-resilient RF fingerprinting,'' \emph{IEEE Communications Magazine}, vol. 58, no. 10, pp. 66--72, Oct. 2020.

\bibitem{Wanglos2019}
F. Wang, Z. Xu, R. Zhi, J. Chen, and P. Zhang, ``LOS/NLOS channel identification technology based on CNN,'' in \emph{2019 6th NAFOSTED Conference on Information and Computer Science (NICS)}, Hanoi, Vietnam, 2019, pp. 200--203.

\bibitem{Aonoperformance2021}
K. Aono, M. Sawahashi, and N. Kamiya, ``Performance of single-carrier LOS-MIMO using FDE in 3GPP TDL channel models,'' in \emph{2020 International Symposium on Antennas and Propagation (ISAP)}, Osaka, Japan, 2021, pp. 819--820.

\bibitem{Noueva2017}
H. Nourollahi and S. G. Maghrebi, ``Evaluation of cyclic prefix length in OFDM system based for Rayleigh fading channels under different modulation schemes,'' in \emph{2017 IEEE Symposium on Computers and Communications (ISCC)}, Heraklion, Greece, 2017, pp. 164--169.

\bibitem{Nacyc2015} T. Nagashima, T. Murakawa, \emph{et al}., ``Cyclic prefix insertion for all-optical fractional OFDM,'' in \emph{2015 International Conference on Photonics in Switching (PS)}, Florence, Italy, 2015, pp. 79--81.

\end{thebibliography}
\end{document}